\documentclass[prb,amsmath,groupedaddress,amssymb,floatfix,footinbib,bibnotes,longbibliography,superscriptaddress,twocolumn]{revtex4-2}

\usepackage{amsmath,amsfonts,amssymb}
\usepackage{mathrsfs}
\usepackage{graphicx}
\usepackage[english]{babel} 
\usepackage{verbatim}
\usepackage[colorlinks=true,citecolor=blue,linkcolor=magenta]{hyperref}
\usepackage[usenames, dvipsnames]{color}
\usepackage{bbold}
\usepackage{chemformula}

\newcommand{\KITP}{Kavli Institute for Theoretical Physics, University of California, Santa Barbara, CA 93106-4030}
\newcommand{\UMD}{Condensed Matter Theory Center and Joint Quantum Institute, Department of Physics, University of Maryland, College Park, Maryland 20742-4111}

\begin{document}

\title{Symmetry breaking in zero-field two-dimensional electron bilayers}

\author{Tessa Cookmeyer}
\email[]{tcookmeyer@kitp.ucsb.edu}
\affiliation{\KITP}

\author{Sankar Das Sarma}
\affiliation{\UMD}

\begin{abstract}
    We theoretically consider bilayers of two dimensional (2D) electron gases as in semiconductor quantum wells, and investigate possible spontaneous symmetry breaking transitions at low carrier densities driven by interlayer Coulomb interactions.  We use a self-consistent technique implementing mean field truncations of the interacting four-fermion terms, and find a U(1) layer symmetry breaking transition at low carrier densities where the individual layer identities are lost leading to an effective pseudospin XY ferromagnet in the 2D plane.  Our results validate earlier theoretical works using simpler restricted Hartree-Fock techniques, and establish the pseudospin XY ferromagnet as a possible low-density symmetry broken phase of 2D bilayers.
\end{abstract}
\maketitle

\section{Introduction}
Two-dimensional electron gases (2DEGs) that can be created in layered semiconductor devices are a nearly perfect solid state platform for the study of free electrons confined to two-dimensional planes. When placed in a large magnetic field, these devices revealed the novel integer and fractional quantum Hall phases of matter \cite{sarma2008perspectives}. A natural extension of these systems was creating bilayer systems with a small separation between two 2DEGs. The layer index now serves as a pseudospin degree of freedom whose symmetry is broken by the finite separation. In a large magnetic field, these bilayer systems can exhibit distinct fractional quantum Hall states \cite{Alicea2009, manoharan1997interaction,yamada2019fractional} including a $\nu=1$ state with interlayer coherence \cite{sarma2008perspectives}.

However, even without a magnetic field, there is a rich variety of phases to be explored in these systems. In a single layer, it is well-known that the Bloch ferromagnetic instability leads to spontaneous spin polarization as the density of electrons decreases. For the bilayer system, it was originally thought that at even smaller densities spontaneous layer polarization (in addition to the spin polarziation) would occur where all the charge would move to one of the two layers \cite{ruden1991exchange}. Experiments appeared to observe this charge transfer \cite{katayama1994observation,Katayama1995,Patel1996,Ying1995}, but closer examination revealed that a symmetric, $XY$-ordered state, where there is coherence between the layers, is lower in energy and the experiments observed charge transfer due to the presence of an electric field that broke the  layer symmetry \cite{Zheng1997,conti1996electron}. This spin-polarized interlayer-coherent state is robust to disorder and  has a broken $U(1)$ symmetry that implies a Goldstone mode \cite{Stern2000, Abedinpour2007}. 

There is now renewed interest in bilayers due to the ability to fabricate heterostructures of 2D materials. Whether through doping, applying a gate-voltage, or optical pumping, the same ingredients can be present in twisted bilayer devices: low-density interacting spinful electrons with a pseudospin layer index. Although the emphasis in these systems is often the effect of the Moir\'e potential, which can lead to flat bands that allow for interaction effects to dominate, these systems could instead allow for testing of predictions originally made for 2DEGs if the effect of the Moir\'e potential is small. Indeed, the two 2D layers can be separated by hexagonal boron nitride for a similar setup to the aforementioned 2DEGs \cite{shimazaki2020strongly,nguyen2023perfect}. 

These platforms raise the question of whether there are still additional phases that may be realized in bilayer 2DEGs. When the layers have finite width and/or there is hopping between the layers, an antisymmetric spin polarization can be preferred \cite{Radtke1996,Radtke1998,Sarma1994,Reboredo1997} and continued exploration of the original problem revealed a small parameter range where a 3-parameter polarized state exists \cite{hanna2000double}. The prediction of these states, therefore, raises the question of whether we can theoretically explore the parameter space in an unbiased way (instead of comparing the energy of candidate symmetry-broken phases).

In this work, inspired by the self-consistent iterative approaches commonly used to solve Sachdev-Ye-Kitaev (SYK) systems \cite{Sachdev1993,Davison2017,Patel2018,Song2017,Fu2016,Pan2021}, we approach these problems by solving the Hartree-Fock equations self-consistently at all $k$-values simultaneously. The old analyses \cite{Zheng1997,hanna2000double} were done using a restricted Hartree-Fock approach without allowing $k$-dependence. It is possible that exploring this $k$-dependence will present interesting new physics. However, we find that the $k$-dependence is small and only appears in small areas of parameter space. Our results nevertheless provide a theoretical justification for the restricted Hartree-Fock approach used so far in the literature uncritically in order to study the spontaneous interlayer coherence phenomena, and reiterates, within a more general theoretical framework, the possible experimental feasibility of a pseudospin symmetry-broken XY interlayer coherent phase in 2D bilayers.

\section{Theory}

Our starting point is the jellium model for two 2DEGs separated by a distance $d$. As is standard, we assume there is a density of positive charges on the top and bottom layers, $\rho_T$ and $\rho_B$, respectively,  and that the total density of electrons is $\rho_B+\rho_T$. We can write the model in second quantization with the following basis functions 
\begin{equation}
    \phi_{\pmb k,a}(\pmb r) = \frac{1}{\sqrt{A}} e^{-i \pmb k \cdot \pmb r} \sqrt{\delta(z-\tau d/2)}\chi_\sigma
\end{equation}
where $\pmb k$ is a two-dimensional vector in the plane of the 2DEGs, $A$ is the area, $a=(\sigma,\tau)$ is a single index combining the spin, $\sigma \in \{\uparrow,\downarrow\}$, and layer, $\tau \in \{1,-1\}$, indices, and $\delta(z)$ is the Dirac delta function. The spin part of the basis function is given by the spinor $\chi_\sigma$ such that $\chi_\sigma^\dagger \chi_{\sigma'} = \delta_{\sigma \sigma'}$ as usual. The resulting expression for the Hamiltonian density is
\begin{equation}
\begin{aligned}
    \frac{H}{A} &= \frac{1}{A}\sum_{\pmb k,a} \frac{k^2}{2m} n_{\pmb k a}-2\pi d e^2\left(\frac{N_T}{A} - \rho_T\right)\left(\frac{N_B}{A} - \rho_B\right) \\
    &+ \frac{\pi e^2}{A^2} \sum_{\pmb k,\pmb p,\pmb q\ne 0} \sum_{a,b} V_{ab}(|\pmb q|) c_{\pmb k+\pmb q,a}^\dagger c_{\pmb p-\pmb q,b}^\dagger c_{\pmb p,b} c_{\pmb k,a}
\end{aligned}
\end{equation}
where $n_{\pmb k a} = c_{\pmb k a}^\dagger c_{\pmb k a}$ 
 is the electron number operator, $N_\tau =\sum_{\pmb k \sigma} n_{\pmb k (\sigma,\tau)}$ is the total number operator per layer, and
\begin{equation}
    V_{(\sigma,\tau),(\sigma',\tau')}(|\pmb q|) = \frac{e^{-d |\pmb q| |\tau-\tau'|/2}}{|\pmb q|}.
\end{equation}
 In deriving this expression, we regularized the divergent integral $\Lambda = \int_0^\infty dk$, and the terms proportional to $\Lambda$ are zero when the total charge of the system is zero, i.e. $(N_T+N_B)/A = \rho_T+\rho_B$, and as $A\to \infty$.

We now perform a mean-field decoupling. We do not allow the superconducting channel, as the interaction is explicitly repulsive (and we ignore Kohn-Luttinger superconductivity), and we assume $\langle c_{\pmb k,a}^\dagger  c_{\pmb p,b}\rangle =\langle c_{\pmb k,a}^\dagger  c_{\pmb k,b}\rangle\delta_{\pmb k,\pmb p}$. Therefore, the only (potentially) non-zero expectations are $\Theta^{\pmb k}_{ab} = \langle c_{\pmb k a}^\dagger c_{\pmb kb}\rangle$, allowing us to arrive at the equation
\begin{equation}
\begin{aligned}
    \frac{H}{A} &= \frac{1}{A}\sum_{\pmb k} \psi_{\pmb k}^\dagger \begin{pmatrix}
        A_{\pmb k}^{(T)} & \mathcal B_{\pmb k} \\
        \mathcal B^\dagger_{\pmb k} & A_{\pmb k}^{(B)}
\end{pmatrix}\psi_{\pmb k} 
+ \frac{e^2}{2A}\sum_{\pmb k,ab}\Theta_{ba}^{\pmb k}\mathcal D_{ab}^{\pmb k} \\ &+2\pi d e^2\left(\frac{\langle N_T\rangle}{A} \frac{\langle N_B\rangle}{A} - \rho_T \rho_B\right)
\end{aligned}
\end{equation}
where 
\begin{equation}
    \mathcal D_{ab}^{\pmb k} = \frac{2\pi}{A}\sum_{\pmb p\ne \pmb k} V_{ab}(|\pmb p-\pmb k|) \Theta_{ab}^{\pmb p},
\end{equation}
\begin{equation}
       \mathcal B_k = - e^2\begin{pmatrix} \mathcal D_{(\uparrow,B),(\uparrow,T)}^{\pmb k} & \mathcal D_{(\downarrow,B),(\uparrow,T)}^{\pmb k} \\ \mathcal D_{(\uparrow,B),(\downarrow,T)}^{\pmb k} & \mathcal D_{(\downarrow,B),(\downarrow,T)}^{\pmb k}
   \end{pmatrix},
\end{equation}
and $\psi_{\pmb k}^T = (c_{\pmb k,(\uparrow,T)}, c_{\pmb k,(\downarrow,T)},c_{\pmb k,(\uparrow,B)}, c_{\pmb k,(\downarrow,B)})$.  The matrix 
\begin{equation}
\begin{aligned}
    \mathcal A_{\pmb k}^{(\tau)} &= \left(\frac{k^2}{2m} - 2\pi e^2 d q_\tau\right) \begin{pmatrix} 1 & 0 \\ 0 & 1 \end{pmatrix}
   \\
   &-e^2 \begin{pmatrix} \mathcal D_{(\uparrow,\tau),(\uparrow,\tau)}^{\pmb k} & \mathcal D_{(\downarrow,\tau),(\uparrow,\tau)}^{\pmb k} \\ (\mathcal D_{(\downarrow,\tau),(\uparrow,\tau)}^{\pmb k} )^* & \mathcal D_{(\downarrow,\tau),(\downarrow,\tau)}^{\pmb k}
   \end{pmatrix}
\end{aligned}
\end{equation}
with $q_T = \langle N_B\rangle/A - \rho_B$ (and vice-versa) denoting how much charge has moved from the opposite layer. Because of the potential non-zero value of $\Theta_{(\sigma T),(\sigma',B)}^{\pmb k}$, the mean-field decoupling of $N_T N_B/A^2$ is not just $\langle N_T\rangle N_B + \langle N_B\rangle N_T - \langle N_T \rangle \langle N_B\rangle$, but the correction term  is suppressed by an additional factor of $A$ and therefore is negligible in the thermodynamic limit.

Our goal is now to find the self-consistent ground state by finding $\Theta^{\pmb k}_{ab}$ and $\langle N_B\rangle$ that specify a mean-field Hamiltonian whose ground-state has the same value of $\Theta^{\pmb k}_{ab}$ and $\langle N_B\rangle$. Because our system is fermionic, we will always have $\Theta^{\pmb k}_{ab} = 0$ when $|\pmb k|>k_{Fa}$, where the Fermi momentum is allowed to depend on spin and layer. When $|\pmb k|<k_{Fa}$, do we even expect $k$ dependence on the $\Theta^{\pmb k}_{ab}$?

The answer is yes, but it is instructive to see when this $k$ dependence occurs. Because $\mathcal D_{ab}^{\pmb k}$ is evaluated as an integral over $\Theta^{\pmb p}_{ab}$, the only $k$ dependence arises from $V_{ab}(|\pmb p-\pmb k|)$. Therefore, although the magnitude may depend on $\pmb k$, the ratio between all of the $\mathcal D_{(\sigma, \tau),(\sigma',\tau)}^{\pmb k}$ in the same layer will be the same at every $\pmb k$ implying $\Theta^{\pmb k}_{(\sigma, \tau),(\sigma',\tau)}=\Theta_{(\sigma, \tau),(\sigma',\tau)}$ when $|\pmb k| < k_{F,(\sigma,\tau)}$. We can also immediately conclude that a non-$k$-dependent spin-rotation in each layer can be performed to set $\mathcal D^{\pmb k}_{(\downarrow,\tau),(\uparrow,\tau)}=0$. 

However, this argument does not apply when we have interlayer coherences, $\mathcal D_{(\sigma,T),(\sigma',B)}^{\pmb k}\ne 0$. Using the gauge degree of freedom $c_{\pmb k a}\to e^{i\phi}c_{\pmb k a}$, we can make all the $\mathcal D^{\pmb k}_{ab}$ that appear in $\mathcal B_{\pmb k}$ real. Now, focusing on only the $\sigma=\uparrow$ spin degree of freedom and suppressing the $\uparrow$ index, at each $\pmb k$, we will need to diagonalize the matrix
\begin{equation}
    \frac{k^2}{2m} I_{2\times 2} -e^2\begin{pmatrix} \mathcal D_{TT}^{(\pmb k)}+ d q_T & \mathcal D_{BT}^{(\pmb k)} \\
    \mathcal D_{BT}^{(\pmb k)} &  D_{BB}^{(\pmb k)}- d q_T \end{pmatrix}
\end{equation}
since $q_T+q_B=0$. When the two layers have equal density, the symmetry between the layers is unbroken and we expect the $\mathcal D_{\tau \tau'}^{(\pmb k)}$ to be even or odd. When they are odd, there is no interlayer coherence, and when they are even $q_T=0$ and $\mathcal D_{TT}^{\pmb k}=\mathcal D_{BB}^{\pmb k}$. In both cases, there is no $k$-dependence in diagonalizing the matrix once again. Once the layers have different densities, there can therefore be additional $k$ dependence, which we will observe in our results.

\section{Numerics}

To prepare for the numerics, we switch to the continuum through $(1/A)\sum_{\pmb k}=(2\pi)^{-2} \int \int dk_x dk_y$ and we switch to dimensionless variables with the rescaling
\begin{equation}
    \tilde k = \frac{k}{\sqrt{4\pi \rho_T}};\ \ \ \ \tilde {\mathcal D}_{ab}^{\pmb k} = \frac{\mathcal D_{ab}^{\pmb k}}{\sqrt{4\pi \rho_T}};\ \ \ \   \langle \tilde n_\tau\rangle = \frac{\langle n_\tau\rangle}{\rho_T},
\end{equation}
and we set $R=\rho_B/\rho_T$ which satisfies $0\le R\le 1$ without loss of generality. The self-consistent equations we now need to solve are
\begin{equation}\label{eq:ntotnorm}
    1+R = \langle \tilde n_T\rangle + \langle \tilde n_B\rangle
\end{equation}
\begin{equation}
    \tilde {\mathcal D}_{ab}^{\tilde {\pmb k}}= \int_0^\infty \tilde p d \tilde p \int_0^{2\pi}\frac{d\theta}{2\pi} \tilde V_{ab}(|\tilde {\pmb k} - \tilde {\pmb p}|) \Theta^{\tilde {\pmb p}}_{ab}
\end{equation}
\begin{equation}
    \tilde V_{ab}(|\tilde {\pmb q}|) = \frac{ e^{-2\frac{ \tilde d}{r_T}\frac{  (\tau-\tau')}{2}|\tilde {\pmb q}|}}{|\tilde {\pmb q}|}
\end{equation}
\begin{equation}\label{eq:Ham_normalized}
\begin{aligned}
   \tilde H &= \frac{2\tilde d}{r_T^2} (\langle \tilde n_T\rangle \langle \tilde n_B\rangle - R)\\
    +\frac{4}{r_T^2}&\int_0^\infty \tilde k d\tilde k \int_0^{2\pi}\frac{d\theta}{2\pi} \left[\psi_{\tilde {\pmb k}}^\dagger \begin{pmatrix} \tilde {\mathcal A}_{\tilde {\pmb k}}^{(T)} & \tilde {\mathcal B}_{\tilde {\pmb k}} \\
    \tilde {\mathcal B}_{\tilde {\pmb k}}^\dagger & \tilde {\mathcal A}_{\tilde {\pmb k}}^{(B)}\end{pmatrix}\psi_{\tilde {\pmb k} } +r_T\frac{\Theta_{ab}^{\tilde {\pmb k}} \tilde {\mathcal D}_{ba}^{\pmb k}}{2}\right]
\end{aligned}
\end{equation}
where $1/a = me^2$ is the effective Bohr radius, $1/r_T = {a\sqrt{\pi \rho_T}}$ is the Wigner-Seitz radius, $\tilde d = d/a$, and $\tilde H = Ha/(A\rho_T e^2)$. As before, the matrices are
\begin{equation}
       \mathcal B_k = - r_T\begin{pmatrix} \tilde {\mathcal D}_{(\uparrow,B),(\uparrow,T)}^{\pmb k} & \tilde {\mathcal D}_{(\downarrow,B),(\uparrow,T)}^{\pmb k} \\\tilde {\mathcal D}_{(\uparrow,B),(\downarrow,T)}^{\pmb k} &\tilde { \mathcal D}_{(\downarrow,B),(\downarrow,T)}^{\pmb k}
   \end{pmatrix};
\end{equation}
\begin{equation}
\begin{aligned}
    \mathcal A_{\pmb k}^{(\tau)} &= \left[\tilde k^2 +\tau \tilde d (\langle \tilde n_T\rangle-1)\right]\begin{pmatrix} 1 & 0 \\ 0 & 1 \end{pmatrix} 
   \\
   &-r_T\begin{pmatrix} \tilde {\mathcal D}_{(\uparrow,\tau),(\uparrow,\tau)}^{\pmb k} & 0 \\ 0 & \mathcal D_{(\downarrow,\tau),(\downarrow,\tau)}^{\pmb k}
   \end{pmatrix}.
\end{aligned}
\end{equation}
and we compute
\begin{equation}
    \langle \tilde n_\tau\rangle 
    = 2\sum_\sigma \int_{0}^\infty \tilde k d\tilde k \int_0^{2\pi} \frac{d\theta}{2\pi} \Theta^{\tilde {\pmb k}}_{(\sigma,\tau),(\sigma,\tau)} .
\end{equation}
\begin{equation}
    \Theta_{ab}^{\tilde {\pmb k}} = \langle c_{\tilde {\pmb k},a}^\dagger c_{\tilde {\pmb k},b}\rangle = \sum_c S_{\tilde {\pmb k},bc} n_F(\lambda_{\tilde {\pmb k}c}) S_{\tilde {\pmb k},ca}^\dagger 
\end{equation}
where $S_{\tilde {\pmb k},ab}$ and $\lambda_{\tilde {\pmb k},a}$ are the matrix of eigenvectors and the eigenvalues, respectively, of the matrix in the second line of Eq.~\eqref{eq:Ham_normalized}, and $n_F(\epsilon)=[1+e^{\beta(\epsilon-\mu)}]^{-1}$ is the Fermi-Dirac distribution. In this work, we will only be working at $T=0$ implying that $n_F(\epsilon) = \theta_\text{HS}(\mu-\epsilon)$ for $\theta_\text{HS}(x)$ the Heaviside step function.

Our parameter space is three-dimensional specified by $r_T$, $\tilde d$, and $R$.  Following work on SYK-like models \cite{Sachdev1993,Davison2017,Patel2018,Song2017,Fu2016,Pan2021}, we will solve these equations iteratively. We assume that the cylindrical symmetry is not broken so that $\Theta_{ab}^{\tilde {\pmb k}}=\Theta_{ab}^{\tilde k}$ has no dependence on the angle $\theta$. We then discretize the function into evenly spaced points $\tilde k \in \{0,\tilde k_{max}\}$ where $\tilde k_{max}>\sqrt{1+R}$ and $\delta \tilde k$ is the distance between points. The only $\theta$ dependence occurs in evaluating the integral
\begin{equation}
   \tilde V_{ab}'(\tilde k,\tilde p)= \int_0^{2\pi} \frac{d\theta}{2\pi}\tilde V_{ab}\left(\sqrt{\tilde k^2 + \tilde p^2 - 2 \tilde k \tilde p \cos(\theta)}\right).
\end{equation}
When $\tilde k = \tilde p$, this integral actually diverges. However, the integral $\int \tilde p d\tilde p \tilde V_{ab}'(\tilde k,\tilde p) \Theta_{ab}^{\tilde p}$ converges. The contribution to the integral from the values where $|\tilde {\pmb k} - \tilde {\pmb p}|<\epsilon$ is $\int_0^{\epsilon}\tilde d\tilde q \Theta^{\tilde {\pmb q}+\tilde {\pmb k}}=\epsilon \Theta^{\tilde k}$ as $\epsilon \to 0$. Therefore, we can just regularize 
\begin{equation}
    \tilde V_{ab}'(\tilde k,\tilde k) = \int_{\delta \theta}^{2\pi - \delta \theta} \frac{d\theta}{2\pi} \tilde V_{ab}\left(\tilde k \sqrt{2-2\cos(\theta)}\right) +  \frac{\delta \theta }{\delta \tilde k}
\end{equation}
where $\epsilon = \delta \theta \tilde k$ and we take $\delta \theta = \delta \tilde k/\tilde k_{\text{max}}$ so that this procedure only affects one of the discretized points.

At the $m$th iterative step, we have some 
guess for the $\Theta^{\tilde {\pmb k}}_{ab}$ and $\langle\tilde n_T\rangle$, labeled $\Theta^{\tilde {\pmb k}}_{ab,m}$ and $\langle\tilde n_{T,m}\rangle$ respectively, from which we can compute the $\tilde {\mathcal D}_{ab}^{\tilde {\pmb k}}$, diagonalize the matrix in the second line of 
Eq.~\eqref{eq:Ham_normalized}, find the value of $\mu$ needed for $n_F(\epsilon)$ such that Eq.~\eqref{eq:ntotnorm} is satisfied, and recompute the ${\Theta^{\tilde {\pmb k},'}_{ab,m}}$ 
and $\langle \tilde n_{T,m}\rangle'$. If $|\Theta_{ab,m}^{\tilde {\pmb k},'}-\Theta_{ab,m}^{\tilde {\pmb k}}|, |\langle \tilde n_{T,m}\rangle'-\langle \tilde n_{T,m}\rangle|<c_\text{tol}$ for all $a,b \in \{T,B\}$ and $\tilde {\pmb k}$, we have reached convergence. If not,  our guess at the next step is given by
\begin{equation}
    \Theta_{ab,m+1}^{\tilde {\pmb k}} = f {\Theta_{ab,m}^{\tilde {\pmb k},'}} + (1-f) \Theta_{ab,m}^{\tilde {\pmb k}}
\end{equation}
(and similarly for $\langle \tilde n_{T,m+1}\rangle$) where $0\le f \le 1$. We typically use $c_\text{tol}=10^{-5}$ and $f=1/2$. 

The most computationally expensive part of finding the numerical solution is the evaluation of $\tilde V_{ab}'(\tilde k,\tilde p)$. Because of the finite number of possible values of $\tilde k$ for a given $\delta \tilde k$ and $\tilde k_\text{max}$, we can fill an array with the values of the function at every possible input at the start of our computation (and optionally save the results to disk to be loaded later). That array can then be used in lieu of reevaluating the function. Each iteration is then very fast (taking less than a second for $N_k = \tilde k_\text{max} / \delta {\tilde k}=4000$), and the numerics have no issue reaching the very strict definition of convergence after hundreds of iterations.

When the difference in energy between self-consistent solutions is small, we find that our numerics, as is often the case, can get stuck in local minima. Therefore, in addition to performing the numerical optimization in an ``unbiased'' way by providing random initial values for the $\Theta^{\tilde {\pmb k}}_{ab}$, we also separately perform the optimization using the $\Theta^{\tilde {\pmb k}}_{ab}$ corresponding to the states $S_0$, $S_1$, and $S_\xi$, defined below, that are perturbed with uniformly distributed random noise of strength $0.02$ to decrease the chance of getting stuck in the local minimum. For each parameter point, we report the observables derived from the state with the minimum energy resulting from this procedure.

\begin{figure}
    \includegraphics[width=0.45\textwidth]{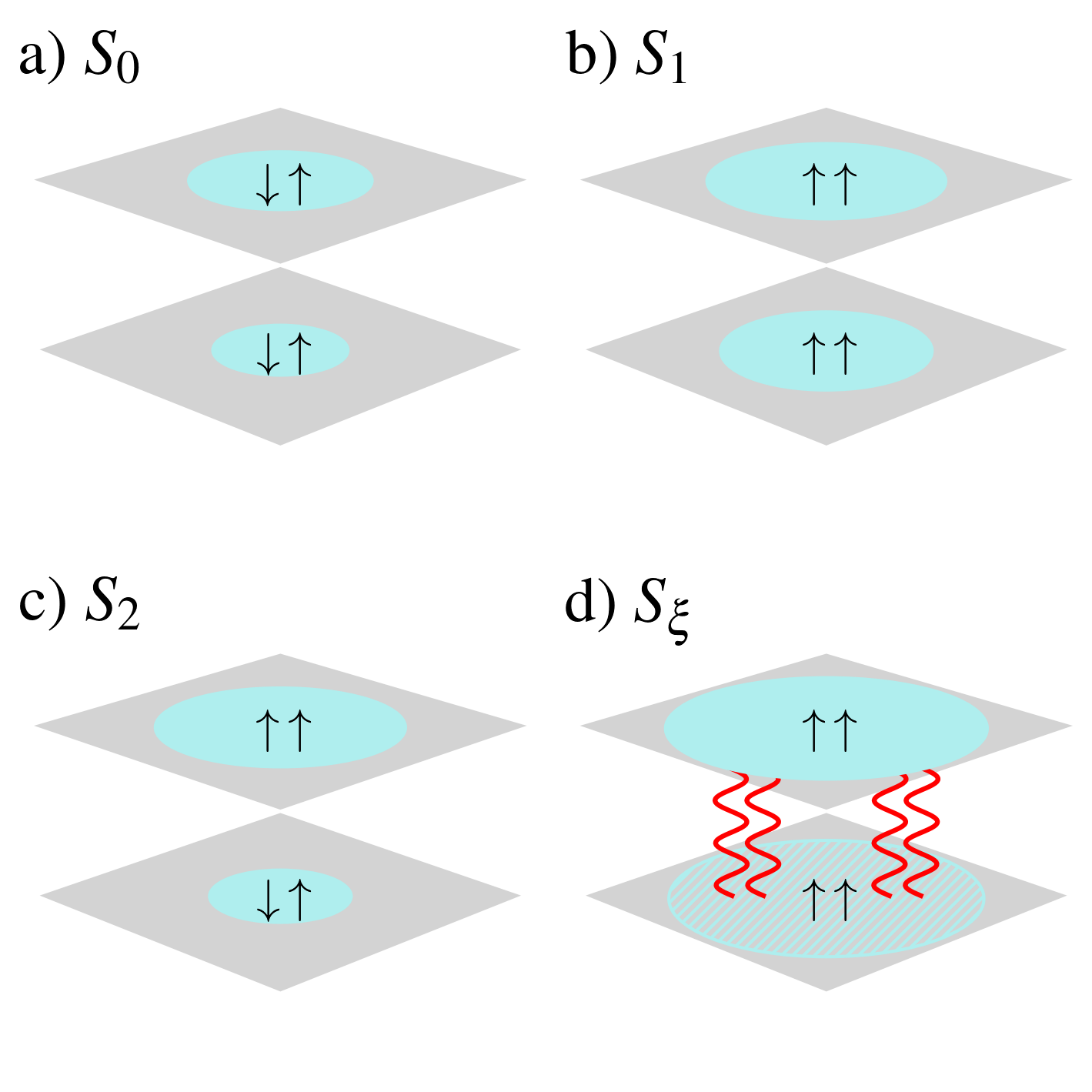}
    \caption{We include cartoon depictions of the trial states $S_0$, $S_1$, $S_2$, and $S_\chi$.  (a) The $S_0$ phase is defined as having no spin polarizations in each layer. (b) The $S_1$ phase occurs when both layers are spin polarized. (c) The $S_2$ phase occurs when one layer is spin-polarized and the other layer is not. In $S_0$, $S_1$, and $S_2$, there is no interlayer coherence and the only variational degree of freedom is how many electrons are in each layer. When the two layers have equal density of positive charges, $R=\rho_B/\rho_T = 1$, the number of electrons in each layer $S_0$ and $S_1$ are the same, but when $R<1$, the two layers can have different numbers of electrons and hence different Fermi surface sizes. In (d) we depict $S_\xi$ where the red lines indicate the interlayer coherence. Here, both layers have the same Fermi surface and the variational degree of freedom is the strength of the coherence,  which is directly related to how many electrons are in each layer. When the coherence is zero, all the electrons are in one layer or the other, which we still define as $S_\xi$. }
    \label{fig:fig0}
\end{figure}

\section{Results and Discussion}
Before discussing our results, we discuss some self-consistent solutions, cartoons of which can be seen in Fig.~\ref{fig:fig0}. In the absence of any interlayer coherences, $\tilde {\mathcal B}_{\tilde {\pmb k}}=0$, all the states can be described by specifying the various densities $\tilde n_{(\sigma,\tau)}$. These states have energy
\begin{equation}\label{eq:nocoh_en}
    \langle \tilde H\rangle = 2\frac{\tilde d}{r_T^2}( \tilde n_{\uparrow T}+  \tilde n_{\downarrow T} -1)^2 + \sum_{a} \frac{n_a^2}{r_T^2} - \frac{8 n_a^{3/2}}{3\pi r_T}
\end{equation}
where $\sum_{a} n_a = 1+R$ and $n_{a} = 2\int_0^{\tilde k_{F,a}} \tilde k d\tilde k$ where $\tilde k_{F,a}$ is the normalized Fermi momentum. We define the state $S_0$ ($S_1$) to be the state that minimizes the above equation and is spin unpolarized (polarized), that is with $\tilde n_{\uparrow\tau}=\tilde n_{\downarrow\tau}=\tilde n_\tau/2$ ($\tilde n_\tau = \tilde n_{\uparrow\tau}$), respectively. These are defined differently than in \cite{Zheng1997} as we perform the minimization over the single free parameter of Eq.~\eqref{eq:nocoh_en} once we include the constraints. Additionally, we define $S_2$ as the minimal energy when one layer is polarized  and the other layer unpolarized \cite{hanna2000double}, which again only requires minimizing one free parameter. It is easy to check that the spins being partially polarized in a single-layer, i.e. $\tilde  n_{\uparrow\tau} \ne \tilde n_{\downarrow \tau}$ with $\tilde  n_{\uparrow\tau}, \tilde n_{\downarrow \tau}\ne 0$, is never energetically favorable, so these states are a complete description of the minima of Eq.~\eqref{eq:nocoh_en}.

We also define $S_\xi$ to be the state with spin polarization and where one of the interlayer coherences is nonzero, e.g. with $\Theta_{(\uparrow,T),(\uparrow,T)}^{\tilde {\pmb k}}=\alpha_T^2\theta_\text{HS}(\tilde k_F-\tilde k)$, $\Theta_{(\uparrow,B),(\uparrow,B)}^{\tilde {\pmb k}}=\alpha_B^2\theta_\text{HS}(\tilde k_F-\tilde k)$  and $\Theta_{(\uparrow,T),(\uparrow,B)}^{\tilde {\pmb k}}=\alpha_T \alpha_B\theta_\text{HS}(\tilde k_F-\tilde k)$ where $\tilde k_F = \sqrt{1+F}$ is the same for both layers and $\alpha_T^2 + \alpha_B^2 = 1$. This state is not necessarily a self-consistent solution to the mean-field equations, but we can still evaluate its energy. We find
\begin{equation}
\begin{aligned}
   \langle \tilde H\rangle &=  \frac{(1+R)^2}{r_T^2}+\frac{2\tilde d}{r_T^2}\left( R \alpha_T^2 - \alpha_B^2\right)^2-  \frac{8(1+R)^{3/2}}{3\pi r_T} \\
   &-  \frac{2(1+R)^{3/2}}{\pi r_T}\alpha_T^2 \alpha_B^2\left[\mathcal I\left(\frac{2 \tilde d}{r_T}\sqrt{1+R}\right) -\frac{8}{3}\right]
\end{aligned}
\end{equation}
with
\begin{equation}
\mathcal I(A) = \int_0^1 s ds \int_0^1 t dt \int_0^{2\pi} d\phi \frac{e^{-A\sqrt{s^2+t^2-2st\cos(\phi)}}}{\sqrt{s^2+t^2-2st\cos(\phi)}}.
\end{equation}
We again have a single parameter minimization  over $\alpha_T$ once we have introduced the constraint. When $R=1$, $S_\xi$ is the interlayer coherent state defined in \cite{Zheng1997}.

\begin{figure*}
    \includegraphics[width=0.95\textwidth]{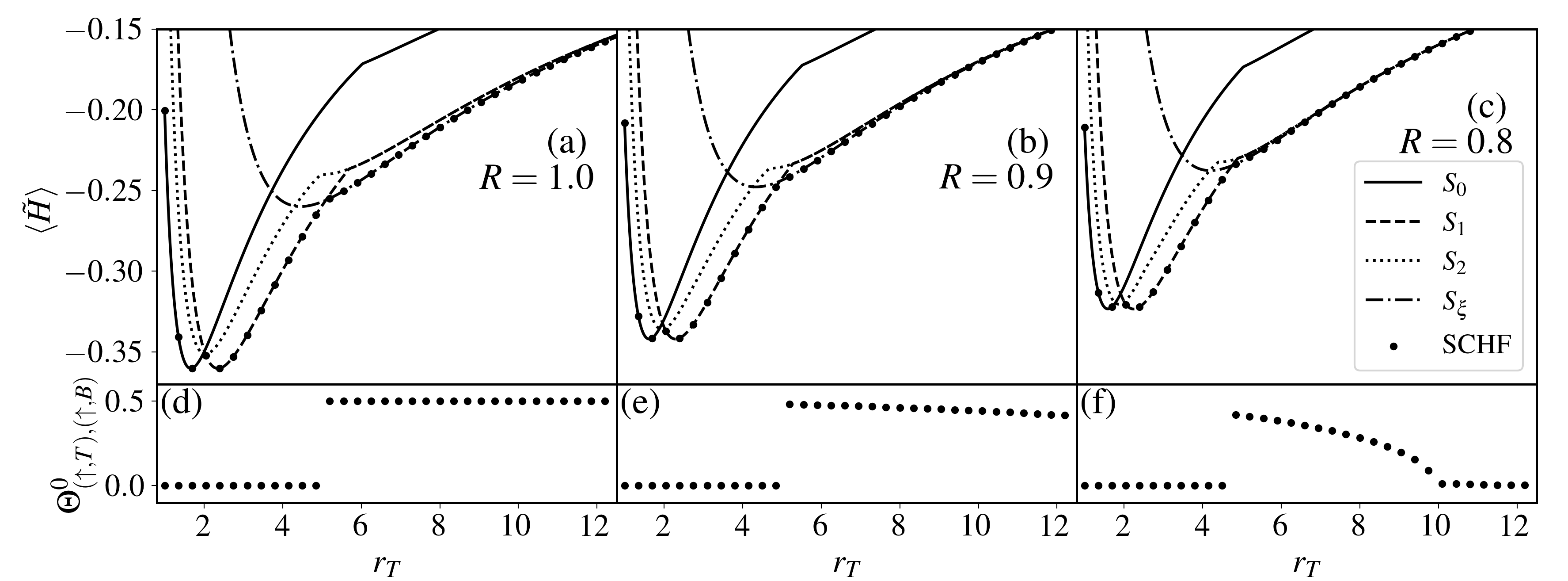}
    \caption{In (a)-(c), we plot the energy, $\langle \tilde H\rangle$ as a function of $r_T$, the dimensionless density parameter, for the trial states $S_0$, $S_1$, $S_2$, and $S_\chi$ as well as our self-consistent solution to the Hartree-Fock equations (SCHF) with the dimensionless distance between layers $\tilde d=1$ ($R=\rho_B/\rho_T$ is the density ratio between the two layers). We accurately reproduce the ground state energy predicted from the four trial states. In (d)-(f), we plot the interlayer coherence parameter $\Theta_{(\uparrow T),(\uparrow, B)}^{0}=\langle c_{0,(\uparrow,T)}^\dagger c_{0,(\uparrow,B)}\rangle $ 
    obtained from our SCHF solution as a function of $r_T$. In the balanced case, $R=1.0$, the interlayer coherence is always present above $r_T\approx 5$, but when $R< 1.0$, there is a maximum $r_T$ for which there is interlayer coherence, beyond which all the electrons have been transferred to the top layer. The value of $R\in\{1.0,0.9,0.8\}$ is the same for the figures in the same column. The numerical parameters $\tilde k_\text{max}=1.5$ and $N_k = \tilde k_\text{max} / \delta \tilde k = 4000$.}
    \label{fig:fig1}
\end{figure*}

\begin{figure}
    \includegraphics[width=0.45\textwidth]{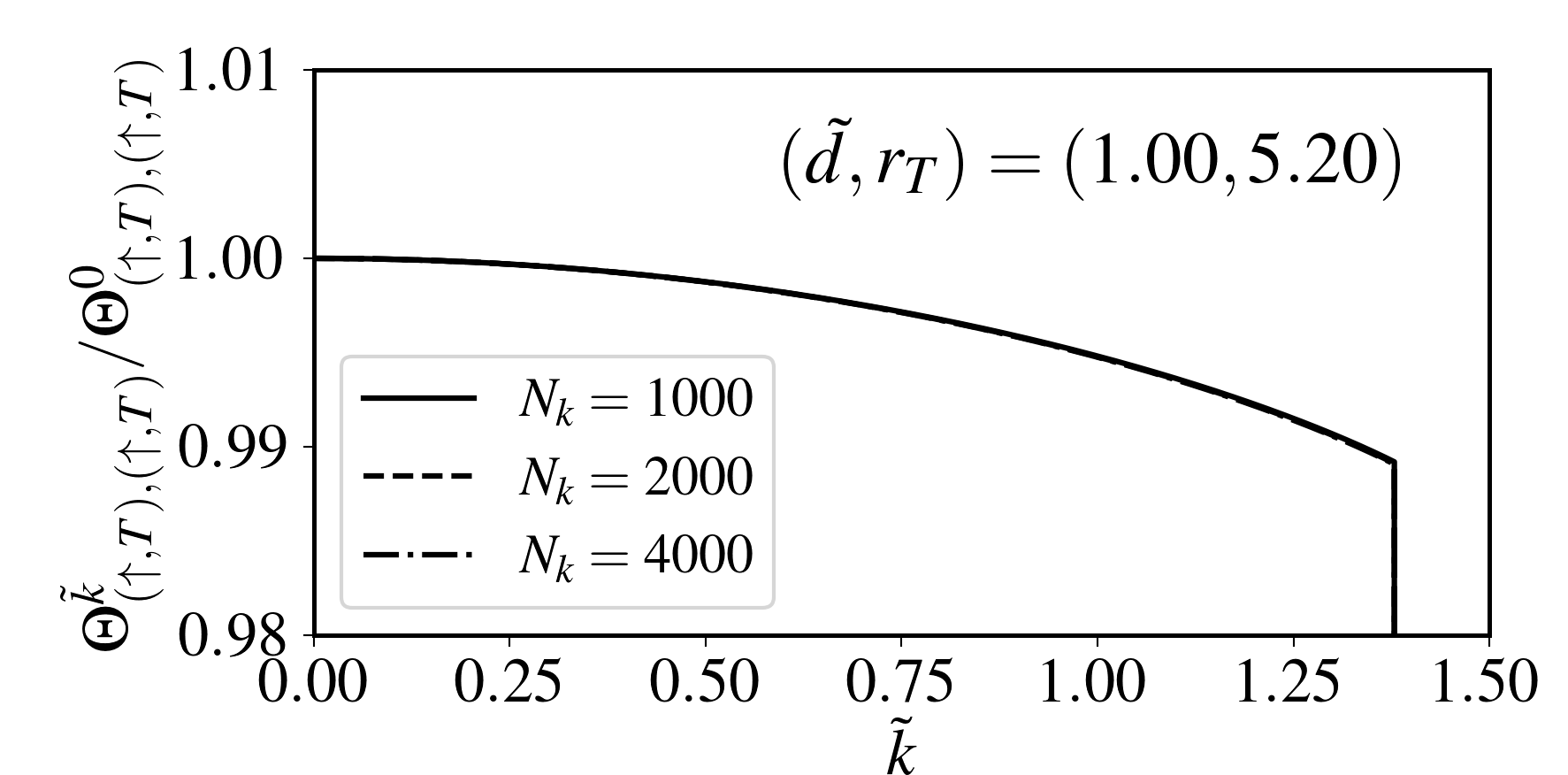}
    \caption{We plot $\Theta_{(\uparrow T,\uparrow T)}^{\tilde k}/\Theta_{(\uparrow T,\uparrow T)}^{0}$ as a function of $\tilde k$ for $\tilde d=1$, $r_T=5.2$, and $R=0.9$. Although there is minimal $k$-dependence, it does not decrease as $N_k = \tilde k_\text{max}/\delta \tilde k$ increases implying that the ground state is slightly different than expected from the state $S_\xi$. However, the energy is only slightly lower: $E_\text{SCHF}(N_k=8000) - E_{S_\xi} =-6\times 10^{-4}$. }
    \label{fig:fig2}
\end{figure}

We compare these energies with those obtained from self-consistently solving the above Hartree-Fock equations (SCHF) in Fig.~\ref{fig:fig1}. We set $\tilde d=1$ as a representative parameter choice. 
In the case of equal density [$R=1$, Fig.~\ref{fig:fig1}(a)] we find that our SCHF approach finds the states $S_0$ (for $r_T< 2$), $S_2$ (for $r_T\approx 2$), $S_1$ (for $2<r_T\lesssim 5$),
and $S_\xi$ (for $r_T\gtrsim 5$) and that the difference in energy between our SCHF approach and the trial states appears to go to zero as the number of
discretized $k$ points, $N_k =\tilde k_\text{max}/ \delta \tilde k$ goes to infinity. It appears that no matter how large $r_T$ gets, the interlayer coherence,  $\Theta^{\tilde k}_{(\uparrow T,\uparrow B)}=\frac12 \theta_\text{HS}(\tilde k_F - \tilde k)$, persists [Fig.~\ref{fig:fig1}(d)] and does not show any 
$\tilde {\pmb k}$ dependence as we predicted from the argument earlier. Particularly, the Ising psudospin ferromagnetic phase, where all electrons spontaneously transfer to one layer at low densities/large $r_T$ (i.e. the state $S_\xi$ where the interlayer coherence is zero) does not appear when $R=1$ as predicted erroneously in earlier work \cite{ruden1991exchange}.  Once we have broken layer symmetry $(R<1)$, we see the strength of the interlayer coherence when $r_T\gtrsim 5$ starts to decrease with increasing $r_T$ eventually goes to essentially zero [Fig.~\ref{fig:fig1}(e)-(f)]. After this occurs, the electrons have been transferred to the top layer.

In the parameter regimes where $R<1$ and interlayer coherence is present, there is some slight $k$-dependence in $\Theta_{(\uparrow,T),(\uparrow,T)}^{\tilde k}$ (Fig.~\ref{fig:fig2}) and therefore also $\Theta_{(\uparrow,B),(\uparrow,B)}^{\tilde k}$ as $\Theta_{(\uparrow,B),(\uparrow,B)}^{\tilde k} + \Theta_{(\uparrow,T),(\uparrow,T)}^{\tilde k}=1$ in this phase. The $k$-dependence is only a slight decrease in the value of $\Theta_{(\uparrow,T),(\uparrow,T)}^{\tilde k}$ away from its $\tilde k=0$ value. As shown in the figure, we have verified that this $k$-dependence is not a finite size effect and is robust to decreasing $\delta \tilde k$. Even in this case, it only slightly lowers the energy $E_\text{SCHF}(N_k=8000) - E_{S_\xi} =-6\times 10^{-4}$ at $(r_T,d,R)=(5.2,1,0.9)$ compared to $E_\text{SCHF}(N_k=8000) - E_{S_\xi} =-5\times 10^{-6}$ at $(r_T,d,R)=(5.2,1,1.0)$, where the latter decrease is due to the finite $N_k$ and is rapidly decreasing with increasing $N_k$. The two necessary ingredients for the $k$-dependence are $R<1$ and interlayer coherence; however, $R<1$ also reduces the parameter space where there  is interlayer coherence implying that the $k$-dependence will never get too large. Thus, our general self-consistent approach validates the restricted Hartree-Fock solutions for the bilayer symmetry breaking, and it appears possible to make the full phase diagram as a function of $(r_T, \tilde d, R)$ using just the trial states we described above, except that the phase boundaries may be slightly shifted from the small decrease in energy. 

\section{Conclusions}
 To conclude, we consider the XY pseudospin symmetry broken spontaneous interlayer coherence in electron bilayers by using a self-consistent mean field approach, generalizing the earlier work using restricted Hartree-Fock theories.  We find that the spontaneous interlayer coherent symmetry-broken XY pseuodoferrmaognetic phase, $S_\xi$ is indeed a possible ground state at low densities (below a critical density).  Although we find that energy of the $S_\xi$ phase can be very slightly lowered by allowing for $k$-dependence in the interlayer coherence (Fig.~\ref{fig:fig2}), the unrestricted nature of our calculation does not reveal any new phases.
 
However, we mention that our self-consistent technique completely settles the slight controversy surrounding the bilayer symmetry breaking phenomenon, where several publications have erroneously claimed that the low-density symmetry-broken ground state might actually be the Ising pseudospin ferromagnetic phase with all the electrons transferring spontaneously to one layer at low densities \cite{Katayama1995,Ying1995,Goni2002,Reboredo1998,Sarma1998}.  Our work establishes beyond any doubt and through a self-consistent theory (transcending the earlier restricted Hartree-Fock theories)  that this Ising ferromagnetic phase always gives way to the xy ferromagnetic phase, $S_{\xi}$, predicted in Ref.~\cite{Zheng1997} where a spontaneous interlayer coherence arises in the system without any charge transfer instability.  Any interlayer charge transfer observed experimentally must thus arise not spontaneously from putative many body phase transition physics, but from some appropriate one electron physics such as the tilting of the wells by explicit electric field effects or through doping one of the wells. Indeed, the Ising pseudospin ferromagnetic phase with all electrons in one layer is preferred at large enough $r_T$ once the layers have unequal electron density $(R<1)$.

\textit{Acknowledgements--} TC is supported by a University of California Presidential Postdoctoral Fellowship and acknowledges support from the Gordon and Betty Moore Foundation through Grant No. GBMF8690 to UCSB. SDS is supported by the Laboratory for Physical Sciences.
Use was made of computational facilities purchased with funds from the National Science Foundation (CNS-1725797) and administered by the Center for Scientific Computing (CSC). The CSC is supported by the California NanoSystems Institute and the Materials Research Science and Engineering Center (MRSEC; NSF DMR 2308708) at UC Santa Barbara.

\end{document}